\documentstyle[twoside,fleqn,espcrc2,epsf]{article}

\title{Geometric Measurement of Topological Susceptibility on Large
Lattices\thanks{Presented by J.~Grandy.  Work supported by
University of California under DOE Contract W-7405-ENG-36.  Computer
time on C90 provided by NERSC and PSC Supercomputing
Centers.}}

\author{J.~Grandy
        and R.~Gupta\address{T-8 Group, MS B285, Los Alamos National
Laboratory, Los Alamos, New Mexico 87545 U.~S.~A.~}}

\begin{document}

\begin{abstract}
The topological susceptibility of the quenched QCD vacuum is measured
on large lattices for three $\beta$ values from $6.0$ to $6.4$.
Charges possibly induced by $O(a)$ dislocations are identified and
shown to have little effect on the measured susceptibility.  As $\beta$
increases, fewer such questionable charges are found.  Scaling is
checked by examining the ratios of the susceptibility to three
physical observables computed in
Refs.~\protect\cite{Schilling,UK62,APE91,Abada} and $\Lambda_{latt}$.
\end{abstract}

\maketitle

The topological susceptibility of the QCD vacuum is defined by 
\begin{equation} \chi_t = \langle Q^2 \rangle / V, \quad
Q = {{-1}\over{16\pi^2}} \int_{V} d^4x \, F(x) \tilde{F}(x) \ .
\end{equation} 
The susceptibility is related to the $\eta'$ and $\eta$ masses
in the chiral large-$N_c$ limit through the Witten-Veneziano
formula\cite{Veneziano81} 
\begin{equation} \chi_t = {{f_\pi^2}\over{6}} (m_{\eta'}^2 + m_\eta^2
- 2m_K^2) = (180 MeV)^4 \ . \end{equation}  

The computation of $\chi_t$ on a lattice is not straightforward.  The
simplest method, replacing $F \tilde{F}$ by a product of plaquettes,
suffers mixing with $F^2$ and a constant\cite{INFN90}, and a large
perturbatively determined renormalization factor, but cooling and
smearing algorithms have been proposed to refine this
method\cite{Teper88,Teper89}.
The geometric method, first proposed by L\"uscher\cite{Luscher}, is
not susceptible to multiplicative renormalization but
contains additive divergences from compact
dislocations\cite{Laursen88} so that
\begin{equation} \chi_t^{measured} = \chi_t + \int_{p_0}^\infty 
dp \, C(p) \, a^p \ . \label{eq:chilatt} \end{equation} 
The dislocation
described by G\"ockeler {\it et al.\/}\cite{Gockeler89} produces a
power $p_0<0$ in (\ref{eq:chilatt}) and thus a power law divergence in
the measured susceptibility as $a\rightarrow 0$.  The factor $C(p)$
for $p < 0$ is crucial to determining the onset of divergent behavior
in the measured susceptibility for finite $a$, and we investigate
whether divergent terms affect the result at realistic values of $a$
used in our calculation.

We are examining several aspects of topology on the lattice.  First,
we compute $\chi_t$ using the geometric method and examine the
systematic effect of dislocations in the range $\beta=6.0$ to
$\beta=6.4$.  We plan to study the various methods of measuring
$\chi_t$, including cooling, and use the geometric topological charge
density to examine the spin asymmetry in deep inelastic scattering off
a proton.  At present we have completed the first stage and report the
results here.

Our lattice ensembles are listed in Table \ref{tab:ensemble}.  We
evaluate the integrals derived by G\"ockeler {\it et
al.\/}\cite{Gockeler87} on the surfaces of each hypercube on the
lattice, obtaining a local integer for each hypercube, and add the
local integers in each configuration to obtain the topological charge
$Q$.  The integration requires a two-stage process.  First, all
hypercubes are integrated using two different coarse ($\sim 5^3$)
cubic meshes.  Hypercubes which show no definite evidence of
convergence toward zero are integrated more accurately, with $\sim
10^3$ meshes on subcubes.  Most hypercubes are brought within
$10^{-4}$ to $10^{-5}$ of an integer with the finer mesh, but a few
difficult hypercubes require further refinement.  The number of
difficult hypercubes per unit topological charge decreases as $\beta$
increases.

\ifx\axisloaded\relax \fi






 
\def\setRevDate $#1 #2 #3${\def\TeXdrawId{TeXdraw V1R3 revised <#2>}}
\setRevDate $Date: Wed Mar  3 12:01:12 MST 1993$

\chardef\catamp=\the\catcode`\@
\catcode`\@=11
\ifx\TeXdraw@included\undefined\global\let\TeXdraw@included=\relax\else
\errhelp{TeXdraw needs to be input only once outside of any groups.}%
\errmessage{Multiple call to include TeXdraw ignored}%
\expandafter \fi

\long                              
\def\centertexdraw #1{\hbox to \hsize{\hss
                                      \btexdraw #1\etexdraw
                                      \hss}}


\def\btexdraw {\x@pix=0             \y@pix=0
               \x@segoffpix=\x@pix  \y@segoffpix=\y@pix
               \t@exdrawdef
               \setbox\t@xdbox=\vbox\bgroup\offinterlineskip
                   \global\d@bs=0           
                   \t@extonlytrue           
                   \p@osinitfalse
                   \savemove \x@pix \y@pix  
                   \m@pendingfalse
                   \p@osinitfalse           
                   \p@athfalse}


\def\etexdraw {\ift@extonly \else
                 \t@drclose      
               \fi
               \egroup           
               \ifdim \wd\t@xdbox>0pt
                 \errmessage{TeXdraw box non-zero size,
                             possible extraneous text}%
               \fi
               \maxhvpos         
               \pixtodim \xminpix \l@lxpos  \pixtodim \yminpix \l@lypos
               \pixtobp {-\xminpix}\l@lxbp  \pixtobp {-\yminpix}\l@lybp
               \vbox {
		      \offinterlineskip
                      \ift@extonly \else
                        \includepsfile{\p@sfile}{\the\l@lxbp}{\the\l@lybp}%
				      {\the\hdrawsize}{\the\vdrawsize}%
                      \fi
                      \vskip\vdrawsize
                      \vskip \l@lypos
                      \hbox {\hskip -\l@lxpos
                             \box\t@xdbox
                             \hskip \hdrawsize
                             \hskip \l@lxpos}%
                      \vskip -\l@lypos\relax}}

%

\def\drawdim #1 {\def\d@dim{#1\relax}}


\def\setunitscale #1 {\edef\u@nitsc{#1}%
                      \realmult \u@nitsc  \s@egsc \d@sc}
\def\relunitscale #1 {\realmult {#1}\u@nitsc \u@nitsc
                      \realmult \u@nitsc \s@egsc \d@sc}
\def\setsegscale #1 {\edef\s@egsc {#1}%
                     \realmult \u@nitsc \s@egsc \d@sc}
\def\relsegscale #1 {\realmult {#1}\s@egsc \s@egsc
                     \realmult \u@nitsc \s@egsc \d@sc}

\def\bsegment {\ifp@ath
                 \flushmove
               \fi
               \begingroup
               \x@segoffpix=\x@pix
               \y@segoffpix=\y@pix
               \setsegscale 1
               \global\advance \d@bs by 1 }
\def\esegment {\endgroup
               \ifnum \d@bs=0
                 \writetx {es}%
               \else
                 \global\advance \d@bs by -1
               \fi}

\def\savecurrpos (#1 #2){\getsympos (#1 #2)\a@rgx\a@rgy
                         \s@etcsn \a@rgx {\the\x@pix}%
                         \s@etcsn \a@rgy {\the\y@pix}}%
\def\savepos (#1 #2)(#3 #4){\getpos (#1 #2)\a@rgx\a@rgy
                            \coordtopix \a@rgx \t@pixa
                            \advance \t@pixa by \x@segoffpix
                            \coordtopix \a@rgy \t@pixb
                            \advance \t@pixb by \y@segoffpix
                            \getsympos (#3 #4)\a@rgx\a@rgy
                            \s@etcsn \a@rgx {\the\t@pixa}%
                            \s@etcsn \a@rgy {\the\t@pixb}}

\def\linewd #1 {\coordtopix {#1}\t@pixa
                \flushbs
                \writetx {\the\t@pixa\space sl}}
\def\setgray #1 {\flushbs
                 \writetx {#1 sg}}
\def\lpatt (#1){\listtopix (#1)\p@ixlist
                \flushbs
                \writetx {[\p@ixlist] sd}}

\def\lvec (#1 #2){\getpos (#1 #2)\a@rgx\a@rgy
                  \s@etpospix \a@rgx \a@rgy
                  \writeps {\the\x@pix\space \the\y@pix\space lv}}
\def\rlvec (#1 #2){\getpos (#1 #2)\a@rgx\a@rgy
                   \r@elpospix \a@rgx \a@rgy
                   \writeps {\the\x@pix\space \the\y@pix\space lv}}
\def\move (#1 #2){\getpos (#1 #2)\a@rgx\a@rgy
                  \s@etpospix \a@rgx \a@rgy
                  \savemove \x@pix \y@pix}
\def\rmove (#1 #2){\getpos (#1 #2)\a@rgx\a@rgy
                   \r@elpospix \a@rgx \a@rgy
                   \savemove \x@pix \y@pix}

\def\lcir r:#1 {\coordtopix {#1}\t@pixa
                \writeps {\the\t@pixa\space cr}%
                \r@elupd \t@pixa \t@pixa
                \r@elupd {-\t@pixa}{-\t@pixa}}
\def\fcir f:#1 r:#2 {\coordtopix {#2}\t@pixa
                     \writeps {#1 \the\t@pixa\space fc}%
                     \r@elupd \t@pixa \t@pixa
                     \r@elupd {-\t@pixa}{-\t@pixa}}
\def\lellip rx:#1 ry:#2 {\coordtopix {#1}\t@pixa
                         \coordtopix {#2}\t@pixb
                         \writeps {\the\t@pixa\space \the\t@pixb\space el}%
                         \r@elupd \t@pixa \t@pixb
                         \r@elupd {-\t@pixa}{-\t@pixb}}
\def\larc r:#1 sd:#2 ed:#3 {\coordtopix {#1}\t@pixa
                            \writeps {\the\t@pixa\space #2 #3 ar}}


\def\ifill f:#1 {\writeps {#1 fl}}     
\def\lfill f:#1 {\writeps {#1 fp}}     



\def\htext #1{\def\testit {#1}%
              \ifx \testit\l@paren
                \let\next=\h@move
              \else
                \let\next=\h@text
              \fi
              \next{#1}}

\def\rtext td:#1 #2{\def\testit {#2}%
                    \ifx \testit\l@paren
                      \let\next=\r@move
                    \else
                      \let\next=\r@text
                    \fi
                    \next td:#1 {#2}}

\def\vtext {\rtext td:90 }

\def\textref h:#1 v:#2 {\ifx #1R%
                          \edef\l@stuff {\hss}\edef\r@stuff {}%
                        \else
                          \ifx #1C%
                            \edef\l@stuff {\hss}\edef\r@stuff {\hss}%
                          \else  
                            \edef\l@stuff {}\edef\r@stuff {\hss}%
                          \fi
                        \fi
                        \ifx #2T%
                          \edef\t@stuff {}\edef\b@stuff {\vss}%
                        \else
                          \ifx #2C%
                            \edef\t@stuff {\vss}\edef\b@stuff {\vss}%
                          \else  
                            \edef\t@stuff {\vss}\edef\b@stuff {}%
                          \fi
                        \fi}

\def\avec (#1 #2){\getpos (#1 #2)\a@rgx\a@rgy
                  \s@etpospix \a@rgx \a@rgy
                  \writeps {\the\x@pix\space \the\y@pix\space (\a@type)
                            \the\a@lenpix\space \the\a@widpix\space av}}

\def\ravec (#1 #2){\getpos (#1 #2)\a@rgx\a@rgy
                   \r@elpospix \a@rgx \a@rgy
                   \writeps {\the\x@pix\space \the\y@pix\space (\a@type)
                             \the\a@lenpix\space \the\a@widpix\space av}}

\def\arrowheadsize l:#1 w:#2 {\coordtopix{#1}\a@lenpix
                              \coordtopix{#2}\a@widpix}
\def\arrowheadtype t:#1 {\edef\a@type{#1}}

\def\clvec (#1 #2)(#3 #4)(#5 #6)%
           {\getpos (#1 #2)\a@rgx\a@rgy
            \coordtopix \a@rgx\t@pixa
            \advance \t@pixa by \x@segoffpix
            \coordtopix \a@rgy\t@pixb
            \advance \t@pixb by \y@segoffpix
            \getpos (#3 #4)\a@rgx\a@rgy
            \coordtopix \a@rgx\t@pixc
            \advance \t@pixc by \x@segoffpix
            \coordtopix \a@rgy\t@pixd
            \advance \t@pixd by \y@segoffpix
            \getpos (#5 #6)\a@rgx\a@rgy
            \s@etpospix \a@rgx \a@rgy
            \writeps {\the\t@pixa\space \the\t@pixb\space 
                      \the\t@pixc\space \the\t@pixd\space 
                      \the\x@pix\space \the\y@pix\space cv}}
\def\e@tendspline#1\endpoints{}
\newtoks\splinet@ks
\def\splinep@int #1 #2 %
           {%
            \advance\p@intnumber by 1\relax
	    \splinet@ks={(#1 #2)}%
            \us@rconvert (#1 #2)\a@rgx\a@rgy
            \futurelet\n@xttok\wr@tesplinepoint}
\def\wr@tesplinepoint{%
            \ifx\n@xttok\endpoints
              \s@etpospix \a@rgx \a@rgy
	      \expandafter
	      \us@rfinish \expandafter
              \p@intnumber \expandafter:\the\splinet@ks=({\x@pix} {\y@pix})%
              \expandafter\e@tendspline
            \else
              \coordtopix \a@rgx\t@pixa
              \advance \t@pixa by \x@segoffpix
              \coordtopix \a@rgy\t@pixb
              \advance \t@pixb by \y@segoffpix
	      \expandafter
              \us@rpoint \expandafter
              \p@intnumber \expandafter:\the\splinet@ks=({\t@pixa} {\t@pixb})%
              \expandafter\splinep@int
            \fi}
\def\defaultus@rfinish#1:(#2 #3)=(#4 #5){\writeps {\the#4 \the#5 \the#1 BSpl}}
\def\defaultus@rpoint#1:(#2 #3)=(#4 #5){\writeps {\the#4 \the#5}}
\def\s@tuseroptions#1/#2/#3/#4@{%
\let\us@rconvert=#1\relax\ifx\us@rconvert\relax\let\us@rconvert=\getpos\fi
\let\us@rpoint=#2\relax\ifx\us@rpoint\relax\let\us@rpoint=\defaultus@rpoint\fi
\let\us@rfinish=#3\relax\ifx\us@rfinish\relax\let\us@rfinish=\defaultus@rfinish
   \fi}%
\def\dooverpoints#1\points{%
    \p@intnumber=0\relax\s@tuseroptions#1///@\splinep@int}
\def\spline{\dooverpoints\points}

\newcount\p@intnumber
\def\curvytype#1{\def\curv@type{#1}}\curvytype{4}%
\def\curvyheight#1{\def\curv@height{#1}}\curvyheight{10}%
\def\curvylength#1{\def\curv@length{#1}}\curvylength{10}%
\def\drawcurvyphoton around (#1 #2) from (#3 #4) to (#5 #6)%
 {\getpos (#1 #2)\a@rgx\a@rgy
  \coordtopix \a@rgx \t@pixa \advance \t@pixa by \x@segoffpix
  \coordtopix \a@rgy \t@pixb \advance \t@pixb by \y@segoffpix
  \writeps {mark \the\t@pixa\space \the\t@pixb}%
  \getpos (#3 #4)\a@rgx\a@rgy
  \s@etpospix \a@rgx\a@rgy
  \writeps {\the\x@pix\space \the\y@pix}%
  \getpos (#5 #6)\a@rgx\a@rgy
  \s@etpospix \a@rgx\a@rgy
  \writeps {\the\x@pix\space \the\y@pix}%
  \writeps {\curv@height\space \curv@length\space \curv@type\space%
              curvyphoton}%
}%
\def\drawcurvygluon around (#1 #2) from (#3 #4) to (#5 #6)%
 {\getpos (#1 #2)\a@rgx\a@rgy
  \coordtopix \a@rgx \t@pixa \advance \t@pixa by \x@segoffpix
  \coordtopix \a@rgy \t@pixb \advance \t@pixb by \y@segoffpix
  \writeps {mark \the\t@pixa\space \the\t@pixb}%
  \getpos (#3 #4)\a@rgx\a@rgy
  \s@etpospix \a@rgx\a@rgy
  \writeps {\the\x@pix\space \the\y@pix}%
  \getpos (#5 #6)\a@rgx\a@rgy
  \s@etpospix \a@rgx\a@rgy
  \writeps {\the\x@pix\space \the\y@pix}%
  \writeps {\curv@height\space 2 mul \curv@length\space \curv@type\space%
              curvygluon}%
}%
\def\blobfreq#1{\def\bl@bfreq{#1}}\blobfreq{0.2}%
\def\blobangle#1{\def\bl@bangle{#1}}\blobangle{0}%
\def\hatchedblob#1{\def\bl@btype{(#1)}}\hatchedblob{B}%
\def\grayblob#1{\def\bl@btype{#1}}%
\def\drawblob xsize:#1 ysize:#2 at (#3 #4)%
{\getpos (#3 #4)\a@rgx\a@rgy \s@etpospix \a@rgx\a@rgy
 \writeps{\the\x@pix\space \the\y@pix}%
 \getpos (#1 #2)\a@rgx\a@rgy 
 \coordtopix \a@rgx\t@pixa \coordtopix\a@rgy\t@pixb \writeps
{\the\t@pixa\space\the\t@pixb\space\bl@bangle\space\bl@bfreq\space\bl@btype}%
 \writeps {blob}%
 \rmove (-#1 -#2)\rmove (#1 #2)\rmove (#1 #2)\rmove (-#1 -#2)}%
\def\drawbb {\bsegment
               \drawdim bp
               \setunitscale 0.24
               \linewd 1           
               \writeps {\the\xminpix\space \the\yminpix\space mv}%
               \writeps {\the\xminpix\space \the\ymaxpix\space lv}%
               \writeps {\the\xmaxpix\space \the\ymaxpix\space lv}%
               \writeps {\the\xmaxpix\space \the\yminpix\space lv}%
               \writeps {\the\xminpix\space \the\yminpix\space lv}%
             \esegment}


\def\getpos (#1 #2)#3#4{\g@etargxy #1 #2 {} \\#3#4%
                        \c@heckast #3%
                        \ifa@st
                          \g@etsympix #3\t@pixa
                          \advance \t@pixa by -\x@segoffpix
                          \pixtocoord \t@pixa #3
                        \fi
                        \c@heckast #4%
                        \ifa@st
                          \g@etsympix #4\t@pixa
                          \advance \t@pixa by -\y@segoffpix
                          \pixtocoord \t@pixa #4
                        \fi}

\def\getsympos (#1 #2)#3#4{\g@etargxy #1 #2 {} \\#3#4%
                           \c@heckast #3%
                           \ifa@st \else
                             \errmessage {TeXdraw: invalid symbolic coordinate}
                           \fi
                           \c@heckast #4%
                           \ifa@st \else
                             \errmessage {TeXdraw: invalid symbolic coordinate}
                           \fi}

\def\listtopix (#1)#2{\def #2{}%
                      \edef\l@ist {#1 }
                      \t@countc=0
                      \loop
                        \expandafter\g@etitem \l@ist \\\a@rgx\l@ist
                        \a@pppix \a@rgx #2
                        \ifx \l@ist\empty
                          \t@countc=1
                        \fi
                      \ifnum \t@countc=0
                      \repeat}


\def\realmult #1#2#3{\dimen0=#1pt
                     \dimen2=#2\dimen0
                     \edef #3{\expandafter\c@lean\the\dimen2}}

\def\intdiv #1#2#3{\t@counta=#1
                   \t@countb=#2
	           \ifnum \t@countb<0
                      \t@counta=-\t@counta
                      \t@countb=-\t@countb
                   \fi
                   \t@countd=1                    
                   \ifnum \t@counta<0
                      \t@counta=-\t@counta
                      \t@countd=-1
                   \fi
	           \t@countc=\t@counta  \divide \t@countc by \t@countb
                   \t@counte=\t@countc  \multiply \t@counte by \t@countb
                   \advance \t@counta by -\t@counte
	           \t@counte=-1
                   \loop
                     \advance \t@counte by 1
	             \ifnum \t@counte<16
                       \multiply \t@countc by 2           
                       \multiply \t@counta by 2           
                       \ifnum \t@counta<\t@countb \else   
                         \advance \t@countc by 1          
                         \advance \t@counta by -\t@countb 
                       \fi
                   \repeat
	           \divide \t@countb by 2         
	           \ifnum \t@counta<\t@countb     
                     \advance \t@countc by 1
                   \fi
                   \ifnum \t@countd<0             
                     \t@countc=-\t@countc
                   \fi
                   \dimen0=\t@countc sp           
                   \edef #3{\expandafter\c@lean\the\dimen0}}

\outer\def\gnewif #1{\count@\escapechar \escapechar\m@ne
  \expandafter\expandafter\expandafter
   \edef\@if #1{true}{\global\let\noexpand#1=\noexpand\iftrue}%
  \expandafter\expandafter\expandafter
   \edef\@if #1{false}{\global\let\noexpand#1=\noexpand\iffalse}%
  \@if#1{false}\escapechar\count@} 
\def\@if #1#2{\csname\expandafter\if@\string#1#2\endcsname}
{\uccode`1=`i \uccode`2=`f \uppercase{\gdef\if@12{}}} 


\def\coordtopix #1#2{\dimen0=#1\d@dim
                     \dimen2=\d@sc\dimen0
                     \t@counta=\dimen2              
                     \t@countb=\s@ppix
                     \divide \t@countb by 2
                     \ifnum \t@counta<0             
                       \advance \t@counta by -\t@countb
                     \else
                       \advance \t@counta by \t@countb
                     \fi
                     \divide \t@counta by \s@ppix
                     #2=\t@counta}

\def\pixtocoord #1#2{\t@counta=#1%
                     \multiply \t@counta by \s@ppix
                     \dimen0=\d@sc\d@dim
                     \t@countb=\dimen0
                     \intdiv \t@counta \t@countb #2}

\def\pixtodim #1#2{\t@countb=#1%
                   \multiply \t@countb by \s@ppix
                   #2=\t@countb sp\relax}

\def\pixtobp #1#2{\dimen0=\p@sfactor pt
                  \t@counta=\dimen0
                  \multiply \t@counta by #1%
                  \ifnum \t@counta < 0             
                    \advance \t@counta by -32768
                  \else
                    \advance \t@counta by 32768
                  \fi
                  \divide \t@counta by 65536
                  #2=\t@counta}
                  
\newcount\t@counta    \newcount\t@countb   
\newcount\t@countc    \newcount\t@countd
\newcount\t@counte
\newcount\t@pixa      \newcount\t@pixb     
\newcount\t@pixc      \newcount\t@pixd
\let\l@lxbp=\t@pixa   \let\l@lybp=\t@pixb  
\let\u@rxbp=\t@pixc   \let\u@rybp=\t@pixd

\newdimen\t@xpos      \newdimen\t@ypos
\let\l@lxpos=\t@xpos  \let\l@lypos=\t@ypos

\newcount\xminpix      \newcount\xmaxpix
\newcount\yminpix      \newcount\ymaxpix

\newcount\a@lenpix     \newcount\a@widpix

\newcount\x@pix        \newcount\y@pix
\newcount\x@segoffpix  \newcount\y@segoffpix
\newcount\x@savepix    \newcount\y@savepix

\newcount\s@ppix       

\newcount\d@bs

\newcount\t@xdnum
\global\t@xdnum=0

\newdimen\hdrawsize    \newdimen\vdrawsize

\newbox\t@xdbox

\newwrite\drawfile

\newif\ifm@pending
\newif\ifp@ath
\newif\ifa@st
\gnewif \ift@extonly
\gnewif\ifp@osinit

\def\l@paren{(}
\def\a@st{*}

\catcode`\%=12
  \def\p@b {
\catcode`\%=14
\catcode`\{=12  \catcode`\}=12  \catcode`\u=1 \catcode`\v=2
  \def\l@br u{v  \def\r@br u}v
\catcode `\{=1  \catcode`\}=2   \catcode`\u=11 \catcode`\v=11

{\catcode`\p=12 \catcode`\t=12
 \gdef\c@lean #1pt{#1}}

\def\sppix#1/#2 {\dimen0=1#2 \s@ppix=\dimen0
                 \t@counta=#1%
                 \divide \t@counta by 2
                 \advance \s@ppix by \t@counta
                 \divide \s@ppix by #1
                 \t@counta=\s@ppix
                 \multiply \t@counta by 65536       
                 \advance \t@counta by 32891        
                 \divide \t@counta by 65782         
                 \dimen0=\t@counta sp
                 \edef\p@sfactor {\expandafter\c@lean\the\dimen0}}

\def\g@etargxy #1 #2 #3 #4\\#5#6{\def #5{#1}%
                                 \ifx #5\empty
                                   \g@etargxy #2 #3 #4 \\#5#6
                                 \else
                                   \def #6{#2}%
                                   \def\next {#3}%
                                   \ifx \next\empty \else
                                     \errmessage {TeXdraw: invalid coordinate}%
                                   \fi
                                 \fi}

\def\c@heckast #1{\expandafter
                  \c@heckastll #1\\}
\def\c@heckastll #1#2\\{\def\testit {#1}%
                        \ifx \testit\a@st
                          \a@sttrue
                        \else
                          \a@stfalse
                        \fi}

\def\g@etsympix #1#2{\expandafter
                     \ifx \csname #1\endcsname \relax
                       \errmessage {TeXdraw: undefined symbolic coordinate}%
                     \fi
                     #2=\csname #1\endcsname}

\def\s@etcsn #1#2{\expandafter
                  \xdef\csname#1\endcsname {#2}}

\def\g@etitem #1 #2\\#3#4{\edef #4{#2}\edef #3{#1}}
\def\a@pppix #1#2{\edef\next {#1}%
                  \ifx \next\empty \else
                    \coordtopix {#1}\t@pixa
                    \ifx #2\empty
                      \edef #2{\the\t@pixa}%
                    \else
                      \edef #2{#2 \the\t@pixa}%
                    \fi
                  \fi}

\def\s@etpospix #1#2{\coordtopix {#1}\x@pix
                     \advance \x@pix by \x@segoffpix
                     \coordtopix {#2}\y@pix
                     \advance \y@pix by \y@segoffpix
                     \u@pdateminmax \x@pix \y@pix}

\def\r@elpospix #1#2{\coordtopix {#1}\t@pixa
                     \advance \x@pix by \t@pixa
                     \coordtopix {#2}\t@pixa
                     \advance \y@pix by \t@pixa
                     \u@pdateminmax \x@pix \y@pix}

\def\r@elupd #1#2{\t@counta=\x@pix
                  \advance\t@counta by #1%
                  \t@countb=\y@pix
                  \advance\t@countb by #2%
                  \u@pdateminmax \t@counta \t@countb}

\def\u@pdateminmax #1#2{\ifnum #1>\xmaxpix
                          \global\xmaxpix=#1%
                        \fi
                        \ifnum #1<\xminpix
                          \global\xminpix=#1%
                        \fi
                        \ifnum #2>\ymaxpix
                          \global\ymaxpix=#2%
                        \fi
                        \ifnum #2<\yminpix
                          \global\yminpix=#2%
                        \fi}

\def\maxhvpos {\t@pixa=\xmaxpix
               \advance \t@pixa by -\xminpix
               \pixtodim  \t@pixa {\dimen2}%
               \global\hdrawsize=\dimen2
               \t@pixa=\ymaxpix
               \advance \t@pixa by -\yminpix
               \pixtodim \t@pixa {\dimen2}%
               \global\vdrawsize=\dimen2\relax}

\def\savemove #1#2{\x@savepix=#1\y@savepix=#2%
                   \m@pendingtrue
                   \ifp@osinit \else
                     \p@osinittrue
                     \global\xminpix=\x@savepix \global\yminpix=\y@savepix
                     \global\xmaxpix=\x@savepix \global\ymaxpix=\y@savepix
                   \fi}

\def\flushmove {\p@osinittrue
                \ifm@pending
                  \writetx {\the\x@savepix\space \the\y@savepix\space mv}%
                  \m@pendingfalse
                  \p@athfalse
                \fi}

\def\flushbs {\loop
                \ifnum \d@bs>0
                  \writetx {bs}%
                  \global\advance \d@bs by -1
              \repeat}
               
\def\h@move #1#2 #3)#4{\move (#2 #3)%
                       \h@text {#4}}
\def\h@text #1{\pixtodim \x@pix \t@xpos
               \pixtodim \y@pix \t@ypos
               \vbox to 0pt{\normalbaselines
                            \t@stuff
                            \kern -\t@ypos
                            \hbox to 0pt{\l@stuff
                                         \kern \t@xpos
                                         \hbox {#1}%
                                         \kern -\t@xpos
                                         \r@stuff}%
                            \kern \t@ypos
                            \b@stuff\relax}}

\def\r@move td:#1 #2#3 #4)#5{\move (#3 #4)%
                             \r@text td:#1 {#5}}
\def\r@text td:#1 #2{\pixtodim \x@pix \t@xpos
                     \pixtodim \y@pix \t@ypos
                     \vbox to 0pt{\kern -\t@ypos
                                  \hbox to 0pt{\kern \t@xpos
                                               \rottxt{#1}{#2}%
                                               \hss}%
                                  \vss}}

\def\rottxt #1#2{\rotsclTeX{#1}{1}{1}{\z@sb{#2}}}%
\def\z@sb #1{\vbox to 0pt{\normalbaselines
                          \t@stuff
                          \hbox to 0pt{\l@stuff
                                       \hbox {#1}%
                                       \r@stuff}%
                          \b@stuff}}

\def\t@exdrawdef {\sppix 300/in            
                  \drawdim in              
                  \edef\u@nitsc {1}
                  \setsegscale 1           
                  \arrowheadsize l:0.16 w:0.08
                  \arrowheadtype t:T
                  \textref h:L v:B }


\def\writeps #1{\flushbs
                \flushmove
                \p@athtrue
                \writetx {#1}}
\def\writetx #1{\ift@extonly
                  \t@extonlyfalse
                  \t@dropen
                \fi
                \w@rps {#1}}
\def\w@rps #1{\immediate\write\drawfile {#1}}

\def\t@dropen {%
  \global\advance \t@xdnum by 1
  \ifnum \t@xdnum<10
    \xdef\p@sfile {\jobname.ps\the\t@xdnum}
  \else
    \xdef\p@sfile {\jobname.p\the\t@xdnum}
  \fi
  \immediate\openout\drawfile=\p@sfile
  \w@rps {\p@b PS-Adobe-3.0 EPSF-3.0}%
  \w@rps {\p@p BoundingBox: (atend)}%
  \w@rps {\p@p Title: TeXdraw drawing: \p@sfile}%
  \w@rps {\p@p Pages: 1 1}%
  \w@rps {\p@p Creator: TeXdraw V1R3}%
  \w@rps {\p@p CreationDate: \the\year/\the\month/\the\day}%
  \w@rps {\p@p DocumentSuppliedResources: ProcSet TeXDraw 2.2 2}%
  \w@rps {\p@p DocumentData: Clean7Bit}%
  \w@rps {\p@p EndComments}%
  \w@rps {\p@p BeginDefaults}%
  \w@rps {\p@p PageNeededResources: ProcSet TeXDraw 2.2 2}%
  \w@rps {\p@p EndDefaults}%
  \w@rps {\p@p BeginProlog}%
  \w@rps {\p@p BeginResource: ProcSet TeXDraw 2.2 2 14696 10668}%
  \w@rps {\p@p VMlocation: local}%
  \w@rps {\p@p VMusage: 14696 10668}%
  \w@rps { /product where}%
  \w@rps {  {pop product (ghostscript) eq /setglobal {pop} def} if}%
  \w@rps { /setglobal where}%
  \w@rps {  {pop currentglobal false setglobal} if}%
  \w@rps { /setpacking where}%
  \w@rps {  {pop currentpacking false setpacking} if}%
  \w@rps {29 dict dup begin}%
  \w@rps {62 dict dup begin}%
  \w@rps { /rad 0 def /radx 0 def /rady 0 def /svm matrix def}%
  \w@rps { /hhwid 0 def /hlen 0 def /ah 0 def /tipy 0 def}%
  \w@rps { /tipx 0 def /taily 0 def /tailx 0 def /dx 0 def}%
  \w@rps { /dy 0 def /alen 0 def /blen 0 def}%
  \w@rps { /i 0 def /y1 0 def /x1 0 def /y0 0 def /x0 0 def}%
  \w@rps { /movetoNeeded 0 def}%
  \w@rps { /y3 0 def /x3 0 def /y2 0 def /x2 0 def}%
  \w@rps { /p1y 0 def /p1x 0 def /p2y 0 def /p2x 0 def}%
  \w@rps { /p0y 0 def /p0x 0 def /p3y 0 def /p3x 0 def}%
  \w@rps { /n 0 def /y 0 def /x 0 def}%
  \w@rps { /anglefactor 0 def /elemlength 0 def /excursion 0 def}%
  \w@rps { /endy 0 def /endx 0 def /beginy 0 def /beginx 0 def}%
  \w@rps { /centery 0 def /centerx 0 def /startangle 0 def }%
  \w@rps { /startradius 0 def /endradius 0 def /elemcount 0 def}%
  \w@rps { /smallincrement 0 def /angleincrement 0 def /radiusincrement 0 def}%
  \w@rps { /ifleft false def /ifright false def /iffill false def}%
  \w@rps { /freq 1 def /angle 0 def /yrad 0 def /xrad 0 def /y 0 def /x 0 def}%
  \w@rps { /saved 0 def}%
  \w@rps {end}%
  \w@rps {/dbdef {1 index exch 0 put 0 begin bind end def}}%
  \w@rps {  dup 3 4 index put dup 5 4 index put bind def pop}%
  \w@rps {/bdef {bind def} bind def}%
  \w@rps {/mv {stroke moveto} bdef}%
  \w@rps {/lv {lineto} bdef}%
  \w@rps {/st {currentpoint stroke moveto} bdef}%
  \w@rps {/sl {st setlinewidth} bdef}%
  \w@rps {/sd {st 0 setdash} bdef}%
  \w@rps {/sg {st setgray} bdef}%
  \w@rps {/bs {gsave} bdef /es {stroke grestore} bdef}%
  \w@rps {/cv {curveto} bdef}%
  \w@rps {/cr \l@br 0 begin}%
  \w@rps { gsave /rad exch def currentpoint newpath rad 0 360 arc}%
  \w@rps { stroke grestore end\r@br\space 0 dbdef}%
  \w@rps {/fc \l@br 0 begin}%
  \w@rps { gsave /rad exch def setgray currentpoint newpath}%
  \w@rps { rad 0 360 arc fill grestore end\r@br\space 0 dbdef}%
  \w@rps {/ar {gsave currentpoint newpath 5 2 roll arc stroke grestore} bdef}%
  \w@rps {/el \l@br 0 begin gsave /rady exch def /radx exch def}%
  \w@rps { svm currentmatrix currentpoint translate}%
  \w@rps { radx rady scale newpath 0 0 1 0 360 arc}%
  \w@rps { setmatrix stroke grestore end\r@br\space 0 dbdef}%
  \w@rps {/fl \l@br gsave closepath setgray fill grestore}%
  \w@rps { currentpoint newpath moveto\r@br\space bdef}%
  \w@rps {/fp \l@br gsave closepath setgray fill grestore}%
  \w@rps { currentpoint stroke moveto\r@br\space bdef}%
  \w@rps {/av \l@br 0 begin /hhwid exch 2 div def /hlen exch def}%
  \w@rps { /ah exch def /tipy exch def /tipx exch def}%
  \w@rps { currentpoint /taily exch def /tailx exch def}%
  \w@rps { /dx tipx tailx sub def /dy tipy taily sub def}%
  \w@rps { /alen dx dx mul dy dy mul add sqrt def}%
  \w@rps { /blen alen hlen sub def}%
  \w@rps { gsave tailx taily translate dy dx atan rotate}%
  \w@rps { (V) ah ne {blen 0 gt {blen 0 lineto} if} {alen 0 lineto} ifelse}%
  \w@rps { stroke blen hhwid neg moveto alen 0 lineto blen hhwid lineto}%
  \w@rps { (T) ah eq {closepath} if}%
  \w@rps { (W) ah eq {gsave 1 setgray fill grestore closepath} if}%
  \w@rps { (F) ah eq {fill} {stroke} ifelse}%
  \w@rps { grestore tipx tipy moveto end\r@br\space 0 dbdef}%
  \w@rps {/setupcurvy \l@br 0 begin}%
  \w@rps { dup 0 eq {1 add} if /anglefactor exch def}%
  \w@rps { abs dup 0 eq {1 add} if /elemlength exch def /excursion exch def}%
  \w@rps { /endy exch def /endx exch def}%
  \w@rps { /beginy exch def /beginx exch def}%
  \w@rps { /centery exch def /centerx exch def}%
  \w@rps { cleartomark}%
  \w@rps { /startangle beginy centery sub beginx centerx sub atan def}%
  \w@rps { /startradius beginy centery sub dup mul }%
  \w@rps {              beginx centerx sub dup mul add sqrt def}%
  \w@rps { /endradius endy centery sub dup mul }%
  \w@rps {            endx centerx sub dup mul add sqrt def}%
  \w@rps { endradius startradius sub }%
  \w@rps { endy centery sub endx centerx sub atan }%
  \w@rps { startangle 2 copy le {exch 360 add exch} if sub dup}%
  \w@rps { elemlength startradius endradius add atan dup add}%
  \w@rps { div round abs cvi dup 0 eq {1 add} if}%
  \w@rps { dup /elemcount exch def }%
  \w@rps { div dup anglefactor div dup /smallincrement exch def}%
  \w@rps { sub /angleincrement exch def}%
  \w@rps { elemcount div /radiusincrement exch def}%
  \w@rps { gsave newpath}%
  \w@rps { startangle dup cos startradius mul }%
  \w@rps { centerx add exch }%
  \w@rps { sin startradius mul centery add moveto}%
  \w@rps { end \r@br 0 dbdef}%
  \w@rps {/curvyphoton \l@br 0 begin}%
  \w@rps { setupcurvy}%
  \w@rps { elemcount \l@br /startangle startangle smallincrement add def}%
  \w@rps {            /startradius startradius excursion add def}%
  \w@rps {            startangle dup cos startradius mul }%
  \w@rps {            centerx add exch }%
  \w@rps {            sin startradius mul centery add}%
  \w@rps {	      /excursion excursion neg def}%
  \w@rps {	      /startangle startangle angleincrement add }%
  \w@rps {                        smallincrement sub def}%
  \w@rps {	      /startradius startradius radiusincrement add def}%
  \w@rps {	      startangle dup cos startradius mul }%
  \w@rps {	      centerx add exch }%
  \w@rps {            sin startradius mul centery add}%
  \w@rps {	      /startradius startradius excursion add def}%
  \w@rps {            /startangle startangle smallincrement add def}%
  \w@rps {             startangle dup cos startradius mul }%
  \w@rps {	       centerx add exch }%
  \w@rps {             sin startradius mul centery add curveto\r@br repeat}%
  \w@rps {	       stroke grestore end}%
  \w@rps {	      \r@br 0 dbdef}%
  \w@rps {/curvygluon \l@br 0 begin}%
  \w@rps { setupcurvy /radiusincrement radiusincrement 2 div def}%
  \w@rps { elemcount \l@br startangle angleincrement add dup}%
  \w@rps {            cos startradius mul centerx add exch}%
  \w@rps {            sin startradius mul centery add}%
  \w@rps {            /startradius startradius radiusincrement add}%
  \w@rps {                         excursion sub def}%
  \w@rps {            startangle angleincrement add dup}%
  \w@rps {            cos startradius mul centerx add exch}%
  \w@rps {            sin startradius mul centery add}%
\w@rps{            startangle angleincrement smallincrement add 2 div add dup}%
  \w@rps {            cos startradius mul centerx add exch}%
  \w@rps {            sin startradius mul centery add}%
  \w@rps {	    curveto}%
\w@rps{      /startangle startangle angleincrement smallincrement add add def}%
  \w@rps {            startangle angleincrement sub dup}%
  \w@rps {            cos startradius mul centerx add exch}%
  \w@rps {            sin startradius mul centery add}%
  \w@rps {	    /startradius startradius radiusincrement add}%
  \w@rps {			excursion add def}%
  \w@rps {            startangle angleincrement sub dup}%
  \w@rps {            cos startradius mul centerx add exch}%
  \w@rps {            sin startradius mul centery add}%
  \w@rps {            startangle dup}%
  \w@rps {            cos startradius mul centerx add exch}%
  \w@rps {            sin startradius mul centery add}%
  \w@rps {	    curveto\r@br repeat}%
  \w@rps { stroke grestore end}%
  \w@rps { \r@br 0 dbdef}%
  \w@rps {/blob \l@br}%
  \w@rps {0 begin st gsave}%
  \w@rps {dup type dup}%
  \w@rps {/stringtype eq}%
  \w@rps {\l@br pop 0 get }%
  \w@rps {dup (B) 0 get eq dup 2 index}%
  \w@rps {(L) 0 get eq or /ifleft exch def}%
  \w@rps {exch (R) 0 get eq or /ifright exch def}%
  \w@rps {/iffill false def \r@br}%
  \w@rps {\l@br /ifleft false def}%
  \w@rps {/ifright false def}%
  \w@rps {/booleantype eq }%
  \w@rps {{/iffill exch def}}%
  \w@rps {{setgray /iffill true def} ifelse \r@br}%
  \w@rps {ifelse}%
  \w@rps {/freq exch def}%
  \w@rps {/angle exch def}%
  \w@rps {/yrad  exch def}%
  \w@rps {/xrad  exch def}%
  \w@rps {/y exch def}%
  \w@rps {/x exch def}%
  \w@rps {newpath}%
  \w@rps {svm currentmatrix pop}%
  \w@rps {x y translate 	}%
  \w@rps {angle rotate}%
  \w@rps {xrad yrad scale}%
  \w@rps {0 0 1 0 360 arc}%
  \w@rps {gsave 1 setgray fill grestore}%
  \w@rps {gsave svm setmatrix stroke grestore}%
  \w@rps {gsave iffill {fill} if grestore}%
  \w@rps {clip newpath}%
  \w@rps {gsave }%
  \w@rps {ifleft  \l@br -3 freq 3 { -1 moveto 2 2 rlineto} for}%
  \w@rps {svm setmatrix stroke\r@br if }%
  \w@rps {grestore}%
  \w@rps {ifright \l@br 3 freq neg -3 { -1 moveto -2 2 rlineto} for}%
  \w@rps {svm setmatrix stroke\r@br if}%
  \w@rps {grestore end}%
  \w@rps {\r@br 0 dbdef}%
  \w@rps {/BSpl \l@br}%
  \w@rps { 0 begin}%
  \w@rps { storexyn}%
  \w@rps { currentpoint newpath moveto}%
  \w@rps { n 1 gt \l@br}%
  \w@rps {  0 0 0 0 0 0 1 1 true subspline}%
  \w@rps {  n 2 gt \l@br}%
  \w@rps {   0 0 0 0 1 1 2 2 false subspline}%
  \w@rps {   1 1 n 3 sub \l@br}%
  \w@rps {    /i exch def}%
  \w@rps {    i 1 sub dup i dup i 1 add dup i 2 add dup false subspline}%
  \w@rps {    \r@br for}%
  \w@rps {   n 3 sub dup n 2 sub dup n 1 sub dup 2 copy false subspline}%
  \w@rps {   \r@br if}%
  \w@rps {  n 2 sub dup n 1 sub dup 2 copy 2 copy false subspline}%
  \w@rps {  \r@br if}%
  \w@rps { end}%
  \w@rps { \r@br 0 dbdef}%
  \w@rps {/midpoint \l@br}%
  \w@rps { 0 begin}%
  \w@rps { /y1 exch def}%
  \w@rps { /x1 exch def}%
  \w@rps { /y0 exch def}%
  \w@rps { /x0 exch def}%
  \w@rps { x0 x1 add 2 div}%
  \w@rps { y0 y1 add 2 div}%
  \w@rps { end}%
  \w@rps { \r@br 0 dbdef}%
  \w@rps {/thirdpoint \l@br}%
  \w@rps { 0 begin}%
  \w@rps { /y1 exch def}%
  \w@rps { /x1 exch def}%
  \w@rps { /y0 exch def}%
  \w@rps { /x0 exch def}%
  \w@rps { x0 2 mul x1 add 3 div}%
  \w@rps { y0 2 mul y1 add 3 div}%
  \w@rps { end}%
  \w@rps { \r@br 0 dbdef}%
  \w@rps {/subspline \l@br}%
  \w@rps { 0 begin}%
  \w@rps { /movetoNeeded exch def}%
  \w@rps { y exch get /y3 exch def}%
  \w@rps { x exch get /x3 exch def}%
  \w@rps { y exch get /y2 exch def}%
  \w@rps { x exch get /x2 exch def}%
  \w@rps { y exch get /y1 exch def}%
  \w@rps { x exch get /x1 exch def}%
  \w@rps { y exch get /y0 exch def}%
  \w@rps { x exch get /x0 exch def}%
  \w@rps { x1 y1 x2 y2 thirdpoint}%
  \w@rps { /p1y exch def}%
  \w@rps { /p1x exch def}%
  \w@rps { x2 y2 x1 y1 thirdpoint}%
  \w@rps { /p2y exch def}%
  \w@rps { /p2x exch def}%
  \w@rps { x1 y1 x0 y0 thirdpoint}%
  \w@rps { p1x p1y midpoint}%
  \w@rps { /p0y exch def}%
  \w@rps { /p0x exch def}%
  \w@rps { x2 y2 x3 y3 thirdpoint}%
  \w@rps { p2x p2y midpoint}%
  \w@rps { /p3y exch def}%
  \w@rps { /p3x exch def}%
  \w@rps { movetoNeeded \l@br p0x p0y moveto \r@br if}%
  \w@rps { p1x p1y p2x p2y p3x p3y curveto}%
  \w@rps { end}%
  \w@rps { \r@br 0 dbdef}%
  \w@rps {/storexyn \l@br}%
  \w@rps { 0 begin}%
  \w@rps { /n exch def}%
  \w@rps { /y n array def}%
  \w@rps { /x n array def}%
  \w@rps { n 1 sub -1 0 \l@br}%
  \w@rps {  /i exch def}%
  \w@rps {  y i 3 2 roll put}%
  \w@rps {  x i 3 2 roll put}%
  \w@rps {  \r@br for end}%
  \w@rps { \r@br 0 dbdef}%
  \w@rps {/bop \l@br save 0 begin /saved exch def end}%
  \w@rps { scale setlinecap setlinejoin setlinewidth setdash moveto}%
  \w@rps { \r@br 1 dbdef}%
  \w@rps {/eop {stroke 0 /saved get restore showpage} 1 dbdef}%
  \w@rps {end /defineresource where}%
  \w@rps { {pop mark exch /TeXDraw exch /ProcSet defineresource cleartomark}}%
  \w@rps { {/TeXDraw exch readonly def} ifelse}%
  \w@rps {/setpacking where {pop setpacking} if}%
  \w@rps {/setglobal where {pop setglobal} if}%
  \w@rps {\p@p EndResource}%
  \w@rps {\p@p EndProlog}%
  \w@rps {\p@p Page: 1 1}%
  \w@rps {\p@p PageBoundingBox: (atend)}%
  \w@rps {\p@p BeginPageSetup}%
  \w@rps {/TeXDraw /findresource where}%
  \w@rps { {pop /ProcSet findresource}}%
  \w@rps { {load} ifelse}%
  \w@rps {begin}%
  \w@rps {0 0 [] 0 3 1 1 \p@sfactor\space \p@sfactor\space bop}%
  \w@rps {\p@p EndPageSetup}%
}

\def\t@drclose {%
  \pixtobp \xminpix \l@lxbp  \pixtobp \yminpix \l@lybp
  \pixtobp \xmaxpix \u@rxbp  \pixtobp \ymaxpix \u@rybp
  \w@rps {\p@p PageTrailer}%
  \w@rps {\p@p PageBoundingBox: \the\l@lxbp\space \the\l@lybp\space
                            \the\u@rxbp\space \the\u@rybp}%
  \w@rps {eop end}%
  \w@rps {\p@p Trailer}%
  \w@rps {\p@p BoundingBox: \the\l@lxbp\space \the\l@lybp\space
                            \the\u@rxbp\space \the\u@rybp}%
  \w@rps {\p@p EOF}%
  \closeout\drawfile
}

\catcode`\@=\catamp
\def\dvialwsetup{
\def\includepsfile##1##2##3##4##5{\special{Insert ##1\space%
                                 }}%
\def\rotsclTeX##1##2##3##4{\special{Insert /dev/null do %
                              3 index exch translate cleartomark %
                              matrix currentmatrix aload pop %
                              7 6 roll restore matrix astore %
                              matrix currentmatrix exch setmatrix %
                              0 0 moveto setmatrix %
                              gsave currentpoint 2 copy translate ##1 rotate %
                              ##2 ##3 scale neg exch neg exch translate %
                              save}%
                   ##4%
                   \special{Insert /dev/null do cleartomark restore %
                          currentpoint grestore moveto save}}%
}
\def\dvipssetup{
\def\includepsfile##1##2##3##4##5{\vbox to 0pt{%
                             \vskip##5%
			     \includegraphics{##1}%
                             \vss}}
\def\rotsclTeX##1##2##3##4{%
		       ##4
}
}
\dvipssetup

\expandafter\ifx\csname fonts are loaded\endcsname\relax\else \fi
\immediate\openin0 localfonts.tex
\ifeof0\relax \else\ifx\localfontsloaded\donotdefinethis\else \fi
\font\seventeenrm=cmr17 \font\twelverm=cmr12 \font\tenrm=cmr10 
\font\ninerm=cmr9       \font\eightrm=cmr8   \font\sevenrm=cmr7
\font\sixrm=cmr6        \font\fiverm=cmr5
\font\twentyfourrm=cmr17 at 24pt \font\twentyrm=cmr17 at 20pt
\font\sixteenrm=cmr17 at 16pt \font\fourteenrm=cmr12 at 14pt
\font\twelvei=cmmi12    \font\teni=cmmi10    \font\ninei=cmmi9
\font\eighti=cmmi8      \font\seveni=cmmi7   \font\sixi=cmmi6
\font\fivei=cmmi5
\font\twentyfouri=cmmi12 at 24pt \font\twentyi=cmmi12 at 20pt
\font\sixteeni=cmmi12 at 16pt \font\fourteeni=cmmi12 at 14pt
\font\tensy=cmsy10      \font\ninesy=cmsy9   \font\eightsy=cmsy8
\font\sevensy=cmsy7     \font\sixsy=cmsy6    \font\fivesy=cmsy5
\skewchar\tensy='60 \skewchar\ninesy='60 \skewchar\eightsy='60
\skewchar\sevensy='60 \skewchar\sixsy='60 \skewchar\fivesy='60
\font\tenex=cmex10      \font\nineex=cmex9   \font\eightex=cmex8
\font\sevenex=cmex7
\font\fiveex=cmex7 at 5pt
\font\twelveit=cmti12   \font\tenit=cmti10   \font\nineit=cmti9
\font\eightit=cmti8     \font\sevenit=cmti7
\font\twentyfourit=cmti12 at 24pt \font\twentyit=cmti12 at 20pt
\font\sixteenit=cmti12 at 16pt \font\fourteenit=cmti12 at 14pt
\font\fiveit=cmti7 at 5pt
\font\twelvesl=cmsl12   \font\tensl=cmsl10   \font\ninesl=cmsl9
\font\eightsl=cmsl8
\font\twentyfoursl=cmsl12 at 24pt \font\twentysl=cmsl12 at 20pt
\font\sixteensl=cmsl12 at 16pt \font\fourteensl=cmsl12 at 16pt
\font\sevensl=cmsl8 at 7pt \font\fivesl=cmsl8 at 5pt
\font\twelvett=cmtt12   \font\tentt=cmtt10   \font\ninett=cmtt9
\font\eighttt=cmtt8
\hyphenchar\twelvett=-1 \hyphenchar\tentt=-1 \hyphenchar\ninett=-1
\hyphenchar\eighttt=-1
\font\twentyfourtt=cmtt12 at 24pt \font\twentytt=cmtt12 at 20pt
\font\sixteentt=cmtt12 at 16pt \font\fourteentt=cmtt12 at 16pt
\font\seventt=cmtt8 at 7pt \font\fivett=cmtt8 at 5pt
\hyphenchar\twentyfourtt=-1 \hyphenchar\twentytt=-1
\hyphenchar\sixteentt=-1 \hyphenchar\fourteentt=-1
\hyphenchar\seventt=-1 \hyphenchar\fivett=-1
\font\tenbf=cmb10
\font\seventeenss=cmss17\font\twelvess=cmss12\font\tenss=cmss10
\font\niness=cmss9      \font\eightss=cmss8
\font\twentyfourss=cmss17 at 24pt \font\twentyss=cmss17 at 20pt
\font\sixteenss=cmss17 at 16pt \font\fourteenss=cmss12 at 14pt
\font\sevenss=cmss8 at 7pt \font\fivess=cmss8 at 5pt
\font\tenmib=cmmib10    \font\ninemib=cmmib9 \font\eightmib=cmmib8
\font\sevenmib=cmmib7   \font\sixmib=cmmib6  \font\fivemib=cmmib5
\skewchar\tenmib='177   \skewchar\ninemib='177 \skewchar\eightmib='177
\skewchar\sevenmib='177 \skewchar\sixmib='177  \skewchar\fivemib='177
\font\tenbsy=cmbsy10    \font\ninebsy=cmbsy9 \font\eightbsy=cmbsy8
\font\sevenbsy=cmbsy7   \font\sixbsy=cmbsy6  \font\fivebsy=cmbsy5
\skewchar\tenbsy='60 \skewchar\ninebsy='60 \skewchar\eightbsy='60
\skewchar\sevenbsy='60 \skewchar\sixbsy='60 \skewchar\fivebsy='60
\let\localfontsloaded=\relax
\fi
\newif\ifplain
\plainfalse
\def\loadfonts#1#2{%
 \expandafter\ifx\csname#1rm\endcsname\relax
 \expandafter\global\expandafter\font\csname#1rm\endcsname=cmr10 at#2pt \fi
 \expandafter\ifx\csname#1i\endcsname\relax
 \expandafter\global\expandafter\font\csname#1i\endcsname =cmmi10 at#2pt 
 \skewchar\csname#1i\endcsname ='177 \fi
 \expandafter\ifx\csname#1sy\endcsname\relax
 \expandafter\global\expandafter\font\csname#1sy\endcsname=cmsy10 at#2pt 
 \skewchar\csname#1sy\endcsname= '60 \fi
 \expandafter\ifx\csname#1ex\endcsname\relax
 \expandafter\global\expandafter\font\csname#1ex\endcsname=cmex10 at#2pt \fi
 \expandafter\ifx\csname#1it\endcsname\relax
 \expandafter\global\expandafter\font\csname#1it\endcsname=cmti10 at#2pt \fi
 \expandafter\ifx\csname#1sl\endcsname\relax
 \expandafter\global\expandafter\font\csname#1sl\endcsname=cmsl10 at#2pt \fi
 \expandafter\ifx\csname#1tt\endcsname\relax
 \expandafter\global\expandafter\font\csname#1tt\endcsname=cmtt10 at#2pt 
 \hyphenchar\csname#1tt\endcsname=  -1 \fi
 \expandafter\ifx\csname#1bf\endcsname\relax
 \expandafter\global\expandafter\font\csname#1bf\endcsname=cmb10  at#2pt \fi
 \expandafter\ifx\csname#1ss\endcsname\relax
 \expandafter\global\expandafter\font\csname#1ss\endcsname=cmss10 at#2pt \fi
 \expandafter\ifx\csname#1mib\endcsname\relax
 \expandafter\global\expandafter\font\csname#1mib\endcsname=cmmib10 at #2pt 
 \skewchar\csname#1mib\endcsname='177 \fi
 \expandafter\ifx\csname#1bsy\endcsname\relax
 \expandafter\global\expandafter\font\csname#1bsy\endcsname=cmbsy10 at #2pt 
 \skewchar\csname#1bsy\endcsname= '60 \fi
}

\ifplain 
\expandafter\ifx\csname tenmib\endcsname\relax\global\font\tenmib=cmmib10 \fi
\expandafter\ifx\csname ninemib\endcsname\global\font\ninemib=cmmib10 at9pt\fi
\expandafter\ifx\csname sevenmi\endcsname\global\font\sevenmib=cmmib10 at7pt\fi
\expandafter\ifx\csname fivemib\endcsname\global\font\fivemib=cmmib10 at5pt\fi
\expandafter\ifx\csname tenbsy\endcsname\global\font\tenbsy=cmbsy10 \fi
\expandafter\ifx\csname ninebsy\endcsname\global\font\ninebsy=cmbsy10 at9pt\fi
\expandafter\ifx\csname sevenbs\endcsname\global\font\sevenbsy=cmbsy10 at7pt\fi
\expandafter\ifx\csname fivebsy\endcsname\global\font\fivebsy=cmbsy10 at5pt\fi
\else
\def\tenfonts{%
	\loadfonts{ten}{10}%
	\global\def\tenfonts{}}
\def\ninefonts{%
	\loadfonts{nine}{9}%
	\global\def\ninefonts{}}
\def\sevenfonts{%
	\loadfonts{seven}{7}%
	\global\def\sevenfonts{}}
\def\fivefonts{%
	\loadfonts{five}{5}%
	\global\def\fivefonts{}}
\fi
\def\twelvefonts{%
	\loadfonts{twelve}{12}%
	\global\def\twelvefonts{}}
\def\fourteenfonts{%
	\loadfonts{fourteen}{14}%
	\global\def\fourteenfonts{}}
\def\sixteenfonts{%
	\loadfonts{sixteen}{16}%
	\global\def\sixteenfonts{}}
\def\twentyfonts{%
	\loadfonts{twenty}{20}%
	\global\def\twentyfonts{}}
\def\twentyfourfonts{%
	\loadfonts{twentyfour}{24}%
	\global\def\twentyfourfonts{}}
\def\famset#1#2#3#4#5{%
	\textfont#1\csname#3#2\endcsname
	\scriptfont#1\csname#4#2\endcsname
	\scriptscriptfont#1\csname#5#2\endcsname}
\def\ninepoint{%
	\ifplain\else\ninefonts\sevenfonts\fivefonts\fi
	\famset0{rm}{nine}{seven}{five}%
	\famset1{i} {nine}{seven}{five}%
	\famset2{sy}{nine}{seven}{five}%
	\famset3{ex}{nine}{seven}{five}%
	\famset\itfam{it}{nine}{seven}{five}%
	\famset\slfam{sl}{nine}{seven}{five}%
	\famset\ttfam{tt}{ten}{seven}{five}%
	\famset\bffam{bf}{nine}{seven}{five}%
	\def\rm{\fam0\ninerm}%
	\def\it{\fam\itfam\nineit}%
	\def\sl{\fam\slfam\ninesl}%
	\def\tt{\fam\ttfam\tentt}%
	\def\bf{\famset0{bf}{nine}{seven}{five}%
		\famset1{mib}{nine}{seven}{five}%
		\famset2{bsy}{nine}{seven}{five}%
		\fam\bffam\ninebf}%
	\setbox\strutbox=\hbox{\vrule height 8pt depth 3pt width 0pt}%
	\baselineskip11pt\rm%
	}
\def\tenpoint{%
	\ifplain\else\tenfonts\sevenfonts\fivefonts\fi
	\famset0{rm}{ten}{seven}{five}%
	\famset1{i} {ten}{seven}{five}%
	\famset2{sy}{ten}{seven}{five}%
	\famset3{ex}{ten}{seven}{five}%
	\famset\itfam{it}{ten}{seven}{five}%
	\famset\slfam{sl}{ten}{seven}{five}%
	\famset\ttfam{tt}{ten}{seven}{five}%
	\famset\bffam{bf}{ten}{seven}{five}%
	\def\rm{\fam0\tenrm}%
	\def\it{\fam\itfam\tenit}%
	\def\sl{\fam\slfam\tensl}%
	\def\tt{\fam\ttfam\tentt}%
	\def\bf{\famset0{bf}{ten}{seven}{five}%
		\famset1{mib}{ten}{seven}{five}%
		\famset2{bsy}{ten}{seven}{five}%
		\fam\bffam\tenbf}%
	\setbox\strutbox=\hbox{\vrule height 8.5pt depth 3.5pt width 0pt}%
	\baselineskip12pt\rm%
	}
\def\twelvepoint{%
	\twelvefonts\ifplain\else\ninefonts\sevenfonts\fi
	\famset0{rm}{twelve}{nine}{seven}%
	\famset1{i} {twelve}{nine}{seven}%
	\famset2{sy}{twelve}{nine}{seven}%
	\famset3{ex}{twelve}{nine}{seven}%
	\famset\itfam{it}{twelve}{nine}{seven}%
	\famset\slfam{sl}{twelve}{nine}{seven}%
	\famset\ttfam{tt}{twelve}{nine}{seven}%
	\famset\bffam{bf}{twelve}{nine}{seven}%
	\def\rm{\fam0\twelverm}%
	\def\it{\fam\itfam\twelveit}%
	\def\sl{\fam\slfam\twelvesl}%
	\def\tt{\fam\ttfam\twelvett}%
	\def\bf{\famset0{bf}{twelve}{nine}{seven}%
		\famset1{mib}{twelve}{nine}{seven}%
		\famset2{bsy}{twelve}{nine}{seven}%
		\fam\bffam\twelvebf}%
	\setbox\strutbox=\hbox{\vrule height 10pt depth 4pt width 0pt}%
	\baselineskip14pt\rm%
	}
\def\fourteenpoint{%
	\fourteenfonts\twelvefonts\ifplain\else\tenfonts\fi
	\famset0{rm}{fourteen}{twelve}{ten}%
	\famset1{i} {fourteen}{twelve}{ten}%
	\famset2{sy}{fourteen}{twelve}{ten}%
	\famset3{ex}{fourteen}{twelve}{ten}%
	\famset\itfam{it}{fourteen}{twelve}{ten}%
	\famset\slfam{sl}{fourteen}{twelve}{ten}%
	\famset\ttfam{tt}{fourteen}{twelve}{ten}%
	\famset\bffam{bf}{fourteen}{twelve}{ten}%
	\def\rm{\fam0\fourteenrm}%
	\def\it{\fam\itfam\fourteenit}%
	\def\sl{\fam\slfam\fourteensl}%
	\def\tt{\fam\ttfam\fourteentt}%
	\def\bf{\famset0{bf}{fourteen}{twelve}{ten}%
		\famset1{mib}{fourteen}{twelve}{ten}%
		\famset2{bsy}{fourteen}{twelve}{ten}%
		\fam\bffam\fourteenbf}%
	\setbox\strutbox=\hbox{\vrule height 12pt depth 5pt width 0pt}%
	\baselineskip17pt\rm%
	}
\def\sixteenpoint{%
	\sixteenfonts\fourteenfonts\twelvefonts
	\famset0{rm}{sixteen}{fourteen}{twelve}%
	\famset1{i} {sixteen}{fourteen}{twelve}%
	\famset2{sy}{sixteen}{fourteen}{twelve}%
	\famset3{ex}{sixteen}{fourteen}{twelve}%
	\famset\itfam{it}{sixteen}{fourteen}{twelve}%
	\famset\slfam{sl}{sixteen}{fourteen}{twelve}%
	\famset\ttfam{tt}{sixteen}{fourteen}{twelve}%
	\famset\bffam{bf}{sixteen}{fourteen}{twelve}%
	\def\rm{\fam0\sixteenrm}%
	\def\it{\fam\itfam\sixteenit}%
	\def\sl{\fam\slfam\sixteensl}%
	\def\tt{\fam\ttfam\sixteentt}%
	\def\bf{\famset0{bf}{sixteen}{fourteen}{twelve}%
		\famset1{mib}{sixteen}{fourteen}{twelve}%
		\famset2{bsy}{sixteen}{fourteen}{twelve}%
		\fam\bffam\sixteenbf}%
	\setbox\strutbox=\hbox{\vrule height 14pt depth 6pt width 0pt}%
	\baselineskip20pt\rm%
	}
\def\twentypoint{%
	\twentyfonts\sixteenfonts\fourteenfonts
	\famset0{rm}{twenty}{sixteen}{fourteen}%
	\famset1{i} {twenty}{sixteen}{fourteen}%
	\famset2{sy}{twenty}{sixteen}{fourteen}%
	\famset3{ex}{twenty}{sixteen}{fourteen}%
	\famset\itfam{it}{twenty}{sixteen}{fourteen}%
	\famset\slfam{sl}{twenty}{sixteen}{fourteen}%
	\famset\ttfam{tt}{twenty}{sixteen}{fourteen}%
	\famset\bffam{bf}{twenty}{sixteen}{fourteen}%
	\def\rm{\fam0\twentyrm}%
	\def\it{\fam\itfam\twentyit}%
	\def\sl{\fam\slfam\twentysl}%
	\def\tt{\fam\ttfam\twentytt}%
	\def\bf{\famset0{bf}{twenty}{sixteen}{fourteen}%
		\famset1{mib}{twenty}{sixteen}{fourteen}%
		\famset2{bsy}{twenty}{sixteen}{fourteen}%
		\fam\bffam\twentybf}%
	\setbox\strutbox=\hbox{\vrule height 17pt depth 7pt width 0pt}%
	\baselineskip24pt\rm%
	}
\def\twentyfourpoint{%
	\twentyfourfonts\twentyfonts\sixteenfonts
	\famset0{rm}{twentyfour}{twenty}{sixteen}%
	\famset1{i} {twentyfour}{twenty}{sixteen}%
	\famset2{sy}{twentyfour}{twenty}{sixteen}%
	\famset3{ex}{twentyfour}{twenty}{sixteen}%
	\famset\itfam{it}{twentyfour}{twenty}{sixteen}%
	\famset\slfam{sl}{twentyfour}{twenty}{sixteen}%
	\famset\ttfam{tt}{twentyfour}{twenty}{sixteen}%
	\famset\bffam{bf}{twentyfour}{twenty}{sixteen}%
	\def\rm{\fam0\twentyfourrm}%
	\def\it{\fam\itfam\twentyfourit}%
	\def\sl{\fam\slfam\twentyfoursl}%
	\def\tt{\fam\ttfam\twentyfourtt}%
	\def\bf{\famset0{bf}{twentyfour}{twenty}{sixteen}%
		\famset1{mib}{twentyfour}{twenty}{sixteen}%
		\famset2{bsy}{twentyfour}{twenty}{sixteen}%
		\fam\bffam\twentyfourbf}%
	\setbox\strutbox=\hbox{\vrule height 17pt depth 7pt width 0pt}%
	\baselineskip24pt\rm%
	}
\expandafter\def\csname fonts are loaded\endcsname{}%

\newif\ifintexdraw
\ifx\axisscale\donotdefinethis\def\axisscale{1}\fi
\def\e{\ifintexdraw\immediate\message{Ending figure}%
       \esegment\etexdraw\ifvmode\medskip\fi\intexdrawfalse\else
       \immediate\message{\string\e\space ignored}\fi}
\def\m#1#2{\move (#1 #2)}
\def\w#1{\linewd #1 }
\def\p#1#2{\move (#1 #2) \fcir f:0 r:1}
\def\l#1#2#3#4{\move (#1 #2) \lvec (#3 #4)}
\def\n#1#2{\lvec (#1 #2)}
\def\s#1#2#3#4{\ifintexdraw\immediate\message{\string\s\space ignored}%
               \else\intexdrawtrue
	       \immediate\message{Starting figure}
               \btexdraw
               \drawdim bp %
               \setunitscale 0.08791 %
	       \bsegment
	       \expandafter\expandafter\expandafter
               \relsegscale\expandafter\axisscale\space
               \fi}
\def\TX{\ifmmode\else$\fi\rm}
\def\XT{\ifmmode$\relax\else\fi}
{\catcode`p=12 \catcode`t=12
 \gdef\dimno#1pt{#1}}%
\def\scle#1#2{{{\dimen0=1000pt
                \dimen0=#2\dimen0\relax
                \expandafter\dimen\expandafter0\expandafter=
                \axisscale\dimen0\relax\divide\dimen0 by 1000\relax
                \xdef\myowntemp{\expandafter\dimno\the\dimen0}}
               \setbox0\hbox{\rotsclTeX{0}{\myowntemp}{\myowntemp}{\hbox{#1}}}%
               \ht0=\myowntemp\ht0
	       \dp0=\myowntemp\dp0
	       \wd0=\myowntemp\wd0
               \box0}}
\def\t#1{\textref h:L v:B \htext{$\rm #1$}}
\def\ltx(#1 #2) #3#4{\textref h:L v:C \htext (#1 #2){\scle{#3\XT}{#4}}}
\def\rtx(#1 #2) #3#4{\textref h:R v:C \htext (#1 #2){\scle{#3\XT}{#4}}}
\def\ctx(#1 #2) #3#4{\textref h:C v:C \htext (#1 #2){\scle{#3\XT}{#4}}}
\def\vtx(#1 #2) #3#4{\textref h:C v:T \vtext (#1 #2){\scle{#3\XT}{#4}}}
\def\solid{\lpatt ()}
\def\disconnected{\lpatt (0 100)}
\def\dotted{\lpatt (10 30)}
\def\dotdashed{\lpatt (10 40 100 100)}
\def\shortdashed{\lpatt (100 100)}
\def\longdashed{\lpatt (200 100)}
\def\dashdotdotted{\lpatt (100 100 10 40 10 40)}
\def\f#1{%
\csname#1\endcsname
}
\let\axisloaded=\relax

\begin{figure}[htb]
\def\axisscale{0.7}
\input {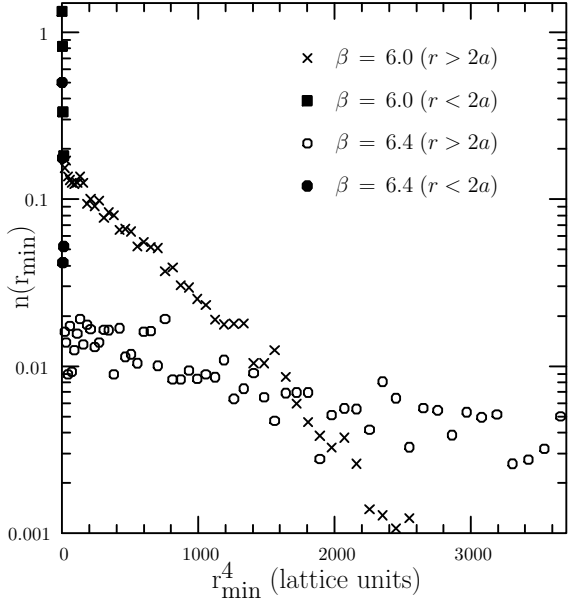}
\vspace{0pt}
\caption{Normalized nearest neighbor correlations
from charged hypercubes to nearest plaquette with $RDC < 0.15$ (Eq.
\protect\ref{eq:RDC}), versus $r_{min}^4$ in lattice units.
Highlighted correlations at $r_{min}<2a$ indicate possible 
compact dislocations. }
\label{fig:dislocs}
\end{figure}

\begin{figure}[t]
\def\axisscale{0.7}
\input {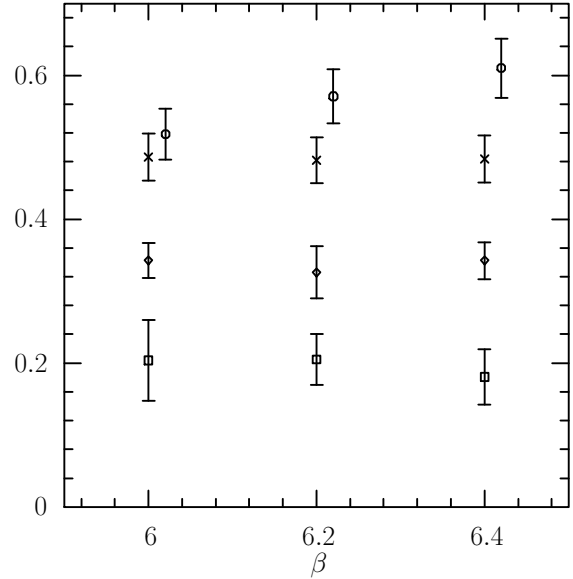}
\vspace{0pt}
\caption{
$\chi_t^{1/4} / \protect\sqrt{\sigma} $ (circles), 
$\chi_t^{1/4} / (100 \Lambda_{\it latt}) $ (crosses), 
$\chi_t^{1/4} / m_\rho $ (diamonds), and 
$\chi_t^{1/4} / (10 f_\pi) $ (squares) versus $\beta$.
Scaling is indicated when the ratios are independent of $\beta$.  The
actual values of these ratios are unimportant, since we consider only
their dependence on $\beta$.  The $16^3 \times 40$ lattice results are
shown for $\beta=6.0$. }
\label{fig:scaling}
\end{figure}

We attempt to separate spurious charges, caused by low-action
dislocations, from physical charges.  These dislocations in $SU(2)$
typically contain a central plaquette near $-1$ and have a radius of
about $1$ lattice spacing\cite{Gockeler89}.  Since the topological
structure of $SU(3)$ is determined by embedded $SU(2)$ windings, we
search for correlations between hypercubes of nonzero charge and
plaquettes near the $SU(3)$ cut locus, the analog to the $-1$ group
element in $SU(2)$.  Plaquettes whose radial distance to cut locus
(RDC)\cite{Lasher89},
\begin{equation} \label{eq:RDC}
RDC = {\left(2 \sum_{i=1}^3 \theta_i^2\right)}^{1/2} \quad 
({{1}\over{\theta_3 - \theta_1}} - {{1}\over{2\pi}}) \end{equation} 
is less than $0.15$ are selected for presentation here.  In this
definition (\ref{eq:RDC}) the $\theta_i$ are the three eigenangles of
the $SU(3)$ matrix in ascending order, so that their sum is zero.  For each
charged hypercube in the ensemble the distance $r_{min}$ from the
center of the hypercube to the center of the nearest selected
plaquette is found.  The number of occurrences at each lattice
distance $r_{min}$, $N(r_{min})$ is divided by the weight factor
$g(r_{min})$, the number of available plaquettes at distance
$r_{min}$, to obtain a normalized nearest neighbor correlation
$n(r_{min})$.  If the charges and nearly critical plaquettes are
decoupled, $n(r_{min})$ will fall exponentially as a function of
volume spanned by $r_{min}$, $n(r_{min})\sim e^{-\alpha r_{min}^4}$.

As we investigate possible dislocations, we find some encouraging
results.  Our plot (Fig. \ref{fig:dislocs}) shows $n(r_{min})$
decaying exponentially as a function of $r_{min}^4$, except for a
dramatic rise for $r_{min} < 2a$, indicating that questionable charges
within $2a$ of a nearly critical plaquette may indeed be dominated by
compact dislocations which are lattice artifacts.  Where $N_q$ is the
number of questionable charges in a configuration, we find that
$\langle N_q \rangle / \langle Q^2\rangle = 0.21$ at $\beta=6.0$, and
$0.05$ at $\beta = 6.4$.  This decrease suggests that for the $\beta$
values considered here the additive term in Eq.~\ref{eq:chilatt} is
dominated by the $p > 0$ part of the integral, and the divergent part
$p<0$ gives at most a negligible contribution to our measured
$\chi_t$.  Also, as $\beta$ increases, the physical charge is spread
over a larger lattice volume and the short range peak of $n(r_{min})$
becomes more pronounced relative to the background exponential,
allowing a cleaner filtering of dislocation-induced charges.  Upon
removing these charges we find that $\chi_t$ changes by about $1/3$ of
the statistical uncertainty for $\beta=6.0$ and only about $1/10$ of
the uncertainty for $\beta=6.4$.  We take this as evidence that
low-action dislocations have little effect on our measured
susceptibility, and present results without removing 
questionable charges.  The decrease of $\langle N_q \rangle / \langle
Q^2\rangle$ and the small effect of questionable charges on measured
$\chi_t$ are also observed for other values of the $RDC$ cutoff.  A
detailed analysis of the $RDC$ cutoff will be presented in a
forthcoming paper\cite{coming}.

\setlength{\tabcolsep}{0.44pc}
\begin{table}[t]
\caption{Lattice Ensembles and Results}
\begin{tabular}{@{\hspace{0.0pc}}rrcrr@{\hspace{0.0pc}}}
\hline 
\multicolumn{1}{c}{$\beta$} & \multicolumn{1}{c}{Lattice} &
\multicolumn{1}{c}{Sample} &
\multicolumn{1}{c}{$\chi_t^{1/4}\left(\sqrt{\sigma}\right)$} &
\multicolumn{1}{c}{$\chi_t^{1/4}\left(m_\rho\right)$}           \\
 & \multicolumn{1}{c}{Size} & \multicolumn{1}{c}{Size} & 
\multicolumn{1}{c}{(MeV)} & \multicolumn{1}{c}{(MeV)}     \\
\hline
$6.0$ & $16^3\times40$ & $34$ & $228(15)\,$ & $264(19)\,$ \\
$6.0$ & $24^3\times40$ & $23$ & $220(29)\,$ & $255(34)\,$ \\
$6.2$ & $32^3\times48$ & $22$ & $251(16)\,$ & $251(28)\,$ \\
$6.4$ & $32^3\times48$ & $21$ & $268(18)\,$ & $264(20)\,$ \\
\hline
\end{tabular}
\label{tab:ensemble}
\end{table}

\setlength{\tabcolsep}{0.7pc}
\begin{table}[b]
\caption{Lattice Observables (Lattice Units)}
\begin{tabular}{@{\hspace{0.0pc}}lrrr@{\hspace{0.0pc}}}
\hline 
\multicolumn{1}{l}{$\beta$} & \multicolumn{1}{c}{$\sigma^{1/2}a$} &
\multicolumn{1}{c}{$m_\rho a$} &
\multicolumn{1}{c}{$f_\pi a$}  \\
\hline
$6.0$ & 0.220(2)\cite{Schilling} & 0.333(8)\cite{APE91} & 
0.056(15)\cite{APE91} \\
$6.2$ & 0.158(1)\cite{Schilling} & 0.277(25)\cite{UK62} & 
0.044(7)\cite{UK62} \\
$6.4$ & 0.119(2)\cite{Schilling} & 0.211(7)\cite{Abada} & 
0.040(8)\cite{Abada} \\
\hline
\end{tabular}
\label{tab:refs}
\end{table}

Since our calculation of $\chi_t$ spans a wide range of $\beta$ we can
check for scaling by plotting dimensionless ratios of $\chi_t^{1/4}$
to four other quantities versus $\beta$.  Our computed values for
$\chi_t^{1/4}a$ are $0.114(8)$ at $\beta=6.0$ ($16^3$ lattice),
$0.110(14)$ at $\beta=6.0$ ($24^3$ lattice),
$0.090(6)$ at $\beta=6.2$, and $0.072(5)$ at $\beta=6.4$.  The height
of each set of points is irrelevant since our concern is whether the
ratios are independent of $\beta$.  Scaling with $m_\rho$, $f_\pi$,
and $\Lambda_{latt}$ derived from the two-loop formula is apparent,
but a possible slope exists in $\chi_t^{1/4} / \sqrt{\sigma}$.  We
present physical values derived from $m_\rho = 770\,{\rm MeV}$ and
$\sqrt{\sigma}=440\,{\rm MeV}$ in Table \ref{tab:ensemble}.  Although
the issue of scaling still must be resolved, we see that these
results are consistent with earlier results by G\"ockeler {\it et
al.\/}\cite{Gockeler87}, computed on much smaller lattices.
Consistency between the smaller and larger lattices at $\beta=6.0$
indicates that our lattices are large enough to control finite size
effects.  We conclude that the geometric method yields $\chi_t$
about a factor of four higher than both the Witten-Veneziano
prediction and the $\chi_t
\approx (180\,{\rm MeV})^4$ obtained by cooling\cite{Teper88}, and that
the superficially divergent part of Eq.~(\ref{eq:chilatt}) is 
unlikely to cause this discrepancy.

\hyphenation{Pa-na-go-pou-los}
 \end{document}